\pdfoutput=1
\pdfmapfile{+stix2.map}

\documentclass[10pt,journal,romanappendices]{IEEEtran}

\usepackage{amsmath,amssymb}
\usepackage{graphicx}
\usepackage{fancyvrb}
\usepackage{xcolor}
\usepackage{framed}
\usepackage{booktabs}
\usepackage{microtype}
\usepackage{cuted,widetext}
\usepackage{cite}
\usepackage{hyperref}

\usepackage{textgreek}
\usepackage[notext]{stix2}

\DeclareMathOperator{\FFT}{\mathcal{F}}
\hypersetup{
    colorlinks,
    linkcolor={red!50!black},
    citecolor={blue!80!black},
    urlcolor={blue!80!black}
}

\title{Counting Statistics of Actively Quenched SPADs Under Continuous Illumination}

\author{
Ivo Straka, Jan Grygar, Josef Hlou\v{s}ek, and Miroslav Je\v{z}ek
\thanks{
This work was supported in part by the Czech Science Foundation under Grant 17-26143S, in part by the National Funding from the MEYS, and in part by the European Union's Horizon 2020 research and innovation framework programme under Grant 731473 (Project 8C18002). Project HYPER-U-P-S has received funding from the QuantERA ERA-NET Cofund in Quantum Technologies implemented within the European Union's Horizon 2020 Programme. The work of J. Grygar and J. Hlou\v{s}ek was supported by Palack\'{y} University under Grant IGA-PrF-2019-010. \emph{(Corresponding author: Ivo Straka.)}

The authors are with the Department of Optics, Faculty of Science, Palack\'{y} University, 17. listopadu, 771~46 Olomouc, Czechia (e-mail: straka@optics.upol.cz; hogr1@seznam.cz; hlousek@optics.upol.cz; jezek@optics.upol.cz)
}
}

\begin{document}

\maketitle

\begin{abstract}
This work presents stochastic approaches to model the counting behavior of actively quenched single-photon avalanche diodes (SPADs) subjected to continuous-wave constant illumination. We present both analytical expressions and simulation algorithms predicting the distribution of the number of detections in a finite time window. We also present formulas for the mean detection rate. The approaches cover recovery time, afterpulsing, and twilight pulsing. We experimentally compare the theoretical predictions to measured data using commercially available silicon SPADs. Their total variation distances range from $\bf 10^{-5}$ to $\bf 10^{-2}$.

\end{abstract}

\section{Single-photon avalanche diodes}

\IEEEPARstart{S}{ingle-photon} avalanche diodes (SPADs) are the most affordable and widespread technology for detecting photons in the field of quantum optics \cite{Migdall2013Book,Chunnilall2014Jul}. The most commonly used materials are silicon for the visible spectrum and InGaAs/InP for the infrared \cite{Buller2007Dec}. SPADs are used either individually for single-photon detection, or integrated -- for some degree of photon-number resolution \cite{Eraerds2007Oct,Chesi2019May}, communications \cite{Zhang2018Feb}, and imaging applications \cite{Bruschini2019Sep}. In particular, multi-pixel SPADs offer new ways to measure the quantum statistics of light \cite{Kalashnikov2012Jul,Chesi2019Mar,Lubin2019Nov}. Multiplexed designs also offer ways of improving the dynamic range of single-photon detection \cite{Antolovic2018Aug,Brida2009Nov}.

This work focuses on actively quenched single SPAD modules that operate on the following principle. An avalanche diode is reverse-biased above breakdown, but its depletion layer contains no free carriers. An incident photon excites an electron-hole pair that causes a rapidly increasing current avalanche. This current is registered by the driving circuit that lowers the bias voltage to stop the avalanche and quench the device. After a pre-set time (dead time), throughout which no other detections can take place, the full voltage is re-applied during a brief reset phase and the SPAD becomes active again.

The SPAD exhibits several properties that are most relevant in photon counting applications: detection efficiency, recovery time, dark counts, afterpulsing, and reset effects \cite{Migdall2013Book}. Efficiency is the probability of registering a detection if a single photon is incident on the SPAD \cite{Polyakov2007Feb,Cohen2018Jul}. Dark counts are detections occurring in the absence of light and commonly arise from spontaneous thermal excitation or tunneling of charge carriers inside the SPAD \cite{Cova1996Apr,Cheng2016Mar}. During the reset phase occurring just after dead time, detections are possible, but they exhibit different conditions including an extra delay due to bias voltage rise time (twilight pulses) \cite{Polyakov2007Feb,Ware2007Jan}. The SPAD therefore exhibits an effective dead time called recovery time, which is the minimum delay between successive detections. Twilight pulses arrive just after the recovery time and their probability of occurrence is proportional to the incident illumination during the rising bias.

Afterpulsing is a spontaneous detection triggered by a released carrier that was trapped in a deep energy level during a previous avalanche \cite{Cova1991Dec}. These traps are mostly caused by material impurities and the probability that a trap captures a carrier during an avalanche is proportional to the avalanche charge. The temporal distribution of afterpulsing depends on the lifetimes of the traps \cite{Humer2015Jul}. This behavior has been treated using various models in the past \cite{Ziarkash2018Mar}.

These effects are a consequence of complex physical processes taking place in the SPAD semiconductor structure and its driving circuitry. Published SPAD models are mainly concerned with a detailed physical description of such processes and simulating the equivalent circuit, which is crucial for designing the SPAD quenching unit \cite{Zappa2009Aug,Cheng2016Mar}. Our aim is to describe the SPAD counting behavior from a user's perspective; that is, model the detection times with respect to incident illumination.

\section{The contribution of this paper}

We present a counting model of the SPAD that determines the number of detections in a time window under continuous-wave illumination \cite{Code}. As the detection process is probabilistic, the output of the model is a probability distribution for the number of detections observed in the time window. The model takes into account all above-stated SPAD imperfections. Additionally, it produces new mean-rate correction formulas. The model is based on simulating a self-exciting point process.  However, suitable approximations are used to derive explicit expressions.

The proposed counting model generalizes known approaches that only consider dead time \cite{Muller1973Sep,Rapp2019May} by incorporating afterpulsing and twilight pulsing. Afterpulsing is also generalized \cite{Kornilov2014Jan,Wang2016Aug}, being treated as a translated point process \cite{Snyder1991}. A Monte Carlo simulation algorithm is provided \cite{Code} that accurately reproduces desired counting statistics under stated assumptions \cite{Stipcevic2013May,Tzou2015Aug}.

The proposed model considers an arbitrary temporal distribution of afterpulses that can be obtained from experimental data; it also considers twilight pulsing, and treats properly the interaction of radiation-induced detections and afterpulses. It assumes a physical model of a continuum of carrier traps \cite{Horoshko2017Jan} with an arbitrary life-time distribution. The model also gives an iterative detection rate formula that accurately incorporates all the discussed phenomena \cite{Code}. Additionally, simplified models of afterpulsing are used to derive explicit relations for counting probabilities and mean detection rates without the need for simulation. These models were already used to verify arbitrarily generated photon statistics in a previous work \cite{Straka2018Apr}.

\section{Point process formulation}

To describe the counting statistics fully, we make a number of assumptions that result in a formulation of a generalized self-exciting point process. This process can be simulated numerically to obtain the counting statistics, and its stationary intensity can be used to calculate the mean event rate.

The basis of the model is a homogeneous Poisson point process with a constant intensity $\mu$ that is modified by considering basic detector imperfections: recovery time $\tau_{\text{R}}$, twilight pulses and afterpulses. The detection efficiency $\eta$ and the dark count rate $\mu_0$ are included in $\mu = \eta \Phi + \mu_0$, where $\Phi$ is the incident photon flux. For more details on the mathematical treatment of point process models here, please refer to Appendix \ref{appendix.point_processes}.

\begin{figure}[t]
\centering
\includegraphics[width=\linewidth]{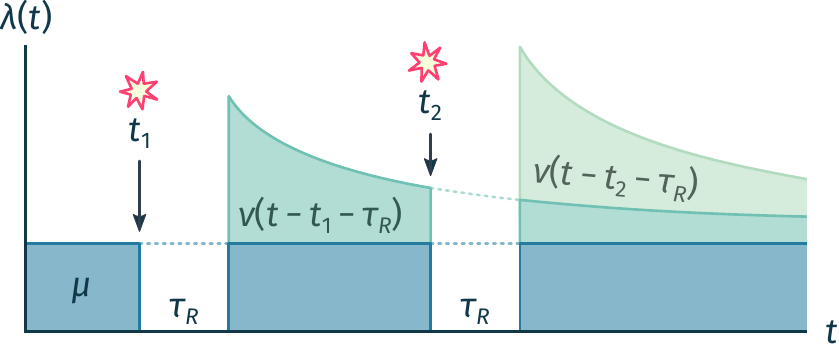}
\caption{An illustration of the self-exciting point process intensity from eq.~\eqref{lambda.pointprocess}. Past detections trigger recovery and excite afterpulses. The scale of the phenomena is exaggerated for illustrative purposes.}
\label{fig.pointprocess}
\end{figure}

Let us first discuss afterpulsing. Conventionally, it is formulated as a single event possibly triggered by each detection with a certain probability, and having a certain temporal probability distribution. The idea is that during each avalanche, there is a chance that a charge carrier gets caught in a deep-level trap and is subsequently released as the level exponentially decays, triggering another avalanche. Previous works suggested that there can be multiple traps with different lifetimes and attempted to model the afterpulse distribution using various mixtures of exponentials, discrete or continuous \cite{Itzler2012Oct,Horoshko2017Jan,Ziarkash2018Mar}.

In this work, we are going to assume a high number of traps, which is supported by the estimated junction volume $\sim 10^{-13}$~m$^3$ and trap concentration $\sim 10^{17}$~m$^{-3}$ \cite{Ghioni2008Feb}. Each type of trap has its own exponential decay rate $\gamma$, its occurrence density in the material and a certain probability of being populated (we assume a constant avalanche charge). These traps work independently \cite{Wang2016Aug}, which means that after each detection, the set of $n_{\text{AP}}$ populated traps $\{\gamma_k\}_{k=1}^{n_{\text{AP}}}$ is a realization of an inhomogeneous Poisson point process with intensity $\rho(\gamma)$. This allows for both discrete and continuous spectra of $\gamma$ (see Appendix \ref{appendix.afterpulsing}). Subsequently, the occurrence of afterpulses in time $t$ becomes a translated point process with intensity $\rho'(t) = \int \gamma \exp(-\gamma t) \rho(\gamma) \mathrm{d}\gamma$. This process is considered unaffected by later detection avalanches \cite{Wang2016Aug} and is therefore superimposed on the counting process of the detector. Because of recovery time, we only consider the offset intensity $\nu(t) \coloneq \rho'(t+\tau_{\text{R}})$. Hence, the number of afterpulses $n_{\text{AP}}$ is a Poisson variable with the mean
\begin{equation}
\langle n_\text{AP} \rangle \coloneq \int_0^\infty \nu(t)\,\mathrm{d}t.
\end{equation}
In the typical limit of $\langle n_\text{AP} \rangle \ll 1$, afterpulsing approaches the model mentioned earlier -- a random choice with a probability $\langle n_\text{AP} \rangle$ and a temporal probability density function (PDF) $p_{\text{AP}}(t) = \nu(t)/\langle n_\text{AP} \rangle$.

Next, we discuss twilight pulses. All carriers accumulated during the SPAD reset (rising bias voltage) are registered approximately at one point just after recovery time \cite{Ware2007Jan}. Due to the reset time being brief, the probability of such an event is approximately proportional to the incident intensity;
\begin{equation}
p_\text{T} \approx \alpha \mu.
\end{equation}
This phenomenon has been called twilight pulsing and has mostly been treated as a linear contribution to afterpulsing. It should be noted that traps decaying during the reset time can also trigger twilight events; however, such events can be included in the afterpulsing intensity $\nu(t)$ without loss of generality and are therefore treated as a part of afterpulsing.

Finally, we consider that during a recovery time $\tau_{\text{R}}$ after each detection, no further events can take place. This enables us to formulate the overall temporal point process, which incorporates everything except twilight pulsing (see also Fig.~\ref{fig.pointprocess}). The process intensity of the $n$-th detection given the history $\{t_i\}$ is

\begin{equation}\label{lambda.pointprocess}
\lambda_n(t_n|\{t_i\}) = 
\begin{cases}
0 & t_n \leq t_{n-1} + \tau_{\text{R}}	\\
\mu + \sum_{i=1}^{n-1} \nu(t_n - t_i - \tau_{\text{R}}) & t_n > t_{n-1} + \tau_{\text{R}}
\end{cases}
.
\end{equation}
This process takes place provided that a twilight pulse did not occur at time $t_{n-1} + \tau_{\text{R}}$ with a probability $p_{\text{T}}$. Taking both possibilities into account, the overall PDF of the $n$-th detection time given $\{t_i\}$ is
\begin{align}\nonumber
p_n(t_n|\{t_i\}) &= p_{\text{T}} \delta(t_n - t_{n-1} - \tau_{\text{R}}) \\\label{pdf.tn}
&+ (1-p_{\text{T}}) \lambda_n(t_n) e^{-\int_{t_{n-1}+\tau_{\text{R}}}^{t_n} \lambda_n(t)\mathrm{d}t}
\end{align}
for $t_n \geq t_{n-1} + \tau_{\text{R}}$ and $\delta$ being the Dirac delta function (derivation of a point process PDF is given in Appendix \ref{appendix.preliminaries}). This process can be numerically simulated using a Monte Carlo approach that is described in Appendix \ref{appendix.simulation} and published on CodeOcean \cite{Code}.

The mean detection rate of such a process can be calculated more efficiently, which is important for practical purposes, such as inferring the real rates $\Phi$ or $\mu$ from the observed detection rate $\mu_\text{det} = \lim_{n \to \infty} n/t_n$. The idea is to find a stationary intensity $\overline\lambda(t)$ and use it to calculate the detection rate.

\begin{figure}[t]
\centering
\includegraphics[width=\linewidth]{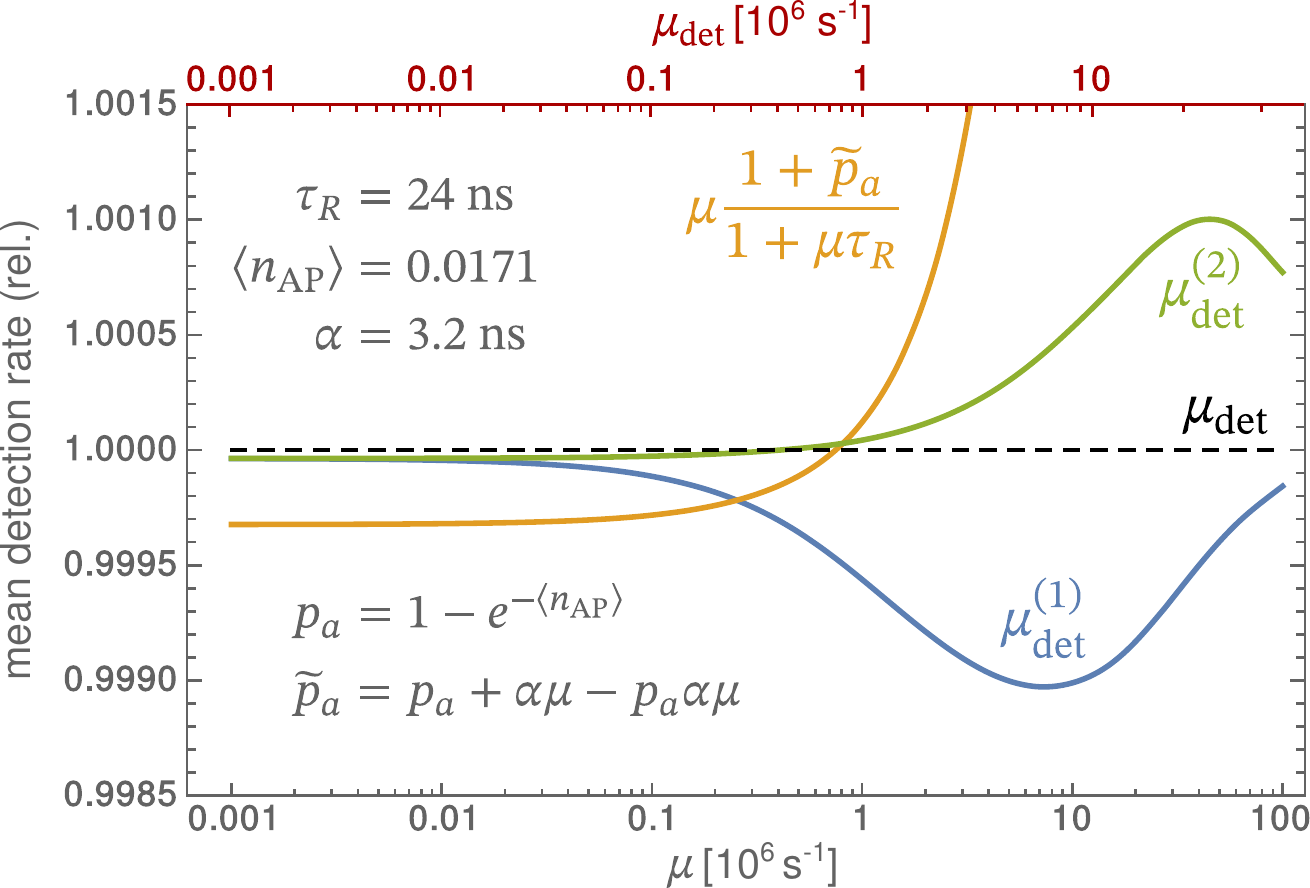}
\caption{The relative accuracy of mean detection rate formulas as a function of the incident Poisson rate $\mu$. The parameters are taken from the detector Excelitas CD3605H due to the most prominent afterpulsing. All values are plotted relatively to the baseline of the most accurate model \eqref{rate.pointprocess}. The orange curve represents a simple multiplication of commonly used correction factors, which becomes highly inaccurate as the detectors approaches saturation. For low rates, the main source of error is the non-binary Poissonian nature of afterpulses. For higher rates, the effect of $\mu_\text{det}^{(1)}$ discarding afterpulses with each new detection leads to underestimating, while $\mu_\text{det}^{(2)}$ overestimates due to counting afterpulses that would happen during recovery.}
\label{fig.meanrates}
\end{figure}

Let us parameterize the time $t_n$ of the $n$-th detection with respect to the most recent recovery: $\Delta t_n \coloneq t_n - t_{n-1} - \tau_{\text{R}} $, $\Delta t_n \in [0,\infty)$. The intensity $\lambda_n$ averaged over its history can be expressed recurrently with $\lambda_1(\Delta t_1) = \mu$ and, using \eqref{pdf.tn},
\begin{align}	\nonumber
\lambda_n(\Delta t_n) &= \int_0^\infty \lambda_{n-1}(\Delta t_{n-1} + \tau_{\text{R}} + \Delta t_n)p_{n-1}(\Delta t_{n-1}) \mathrm{d}\Delta t_{n-1}	\\\label{lambda.stationary}
 &+ \nu(\Delta t_n),
\end{align}
meaning the already existing intensity $\lambda_{n-1}$ is displaced by the delay of the previous detection $\Delta t_{n-1}$, averaged, and afterpulsing $\nu$ is added. The stationary condition is $\overline\lambda = \lambda_n = \lambda_{n-1}$. If we consider only the extra addition $f(\Delta t)$ to the constant intensity, $f(\Delta t) \coloneq \overline\lambda(\Delta t) - \mu$, and substitute in \eqref{lambda.stationary}, we obtain an integral equation for $f$
\begin{align}\nonumber
f(\Delta t) &= (1-p_{\text{T}})\int_0^\infty f(t+\tau_{\text{R}}+\Delta t)(\mu + f(t))e^{-\mu t - F(t)}\mathrm{d}t	\\\label{f.stationary}
 &+ p_{\text{T}} f(\Delta t + \tau_{\text{R}}) + \nu(\Delta t)
\end{align}
with $F(t) \coloneq \int_0^t f(t')$, using the parameters $\mu$, $\tau_{\text{R}}$, $p_{\text{T}}$, and $\nu(t)$. A solution can be found efficiently using an iterative approach, starting with $f_1(t) = 0$ and substituting the left side in the right side repeatedly. Arriving at the solution $\overline\lambda(\Delta t)$, it can be substituted in \eqref{pdf.tn} instead of $\lambda_n$ to get the stationary PDF
\begin{equation}\label{p.stationary}
\overline{p}(\Delta t) = p_{\text{T}} \delta(\Delta t) + (1-p_{\text{T}})\overline{\lambda}(\Delta t) \exp \left(-\int_0^{\Delta t} \overline{\lambda}(t)\mathrm{d}t \right).
\end{equation}
Assuming ergodicity, the mean detection rate is then obtained by averaging,
\begin{equation}\label{rate.pointprocess}
\mu_\text{det} = \left(\langle\Delta t\rangle_{\overline{p}} + \tau_{\text{R}} \right)^{-1}.
\end{equation}
Numerically, the cross-correlation in equation \eqref{f.stationary} can be efficiently evaluated using Fast Fourier Transform. This enables the relation \eqref{rate.pointprocess} to be numerically inverted with respect to $\mu$, which provides detection rate correction using parameters that can be experimentally obtained with time-resolved detection techniques. For details on implementation, see Appendix \ref{appendix.rate} and the CodeOcean capsule \cite{Code}.

\section{Approximate formulations}

We are going to introduce approximations that allow for the counting model to be described analytically. The first step is adopting an interarrival approach, where the whole process is described by a single PDF $p(\Delta t)$. Afterpulsing is considered to be a simple discrete-choice process with a probability $p_a$ and a temporal PDF $p_{\text{AP}}(\Delta t)$. This approach neglects the excitation of multiple afterpulses and their persistence over subsequent avalanches. The probabilistic mixture of all processes leads to
\begin{align}\nonumber
p(\Delta t) &= p_{\text{T}}\delta(\Delta t) + (1-p_{\text{T}})p_a \left(p_{\text{AP}}(\Delta t)-p_{\text{AP}}(\Delta t)\mu \right)e^{-\mu \Delta t}	\\\label{p.iamodel}
& + (1-p_{\text{T}})\mu e^{-\mu \Delta t},
\end{align}
where $p_{\text{AP}}(\Delta t) \coloneq \int_0^{\Delta t}p_{\text{AP}}(t)\mathrm{d}t$. By averaging, we get
\begin{equation}\label{rate.iamodel}
\mu_\text{det}^{(1)} = \left[ \frac{1}{\mu}(1-p_{\text{T}})\left( 1-p_a\int_0^\infty \hspace{-8pt} p_{\text{AP}}(t)e^{-\mu t}\mathrm{d}t\right) + \tau_{\text{R}} \right]^{-1}.
\end{equation}
Here the main element is the integral Laplace transform that eliminates those afterpulses that have not triggered before the next Poissonian event.

The next approximation is to neglect the PDF $p_{\text{AP}}(\Delta t)$ and consider all afterpulses and twilight pulses to arrive together at $\Delta t = 0$ with probability $\widetilde{p}_a$. This final simplification allows us to treat the whole problem analytically and derive expressions for the counting statistics. The mean detection rate \eqref{rate.iamodel} becomes
\begin{equation}\label{rate.simple}
\mu_\text{det}^{(2)} = \frac{\mu}{1-\widetilde{p}_a + \mu \tau_{\text{R}}}.
\end{equation}
\vspace{0pt}

If $\widetilde{p}_a$ is constant or increases with the first power of $\mu$, this formula can be directly inverted. All of the above models of the mean detection rate are compared in Fig.~\ref{fig.meanrates} along with the conventionally used corrections for dead time $\mu_\text{det}^\text{dead} = \mu/(1 + \mu\tau_{\text{R}})$ and afterpulses $\mu_\text{det}^\text{AP} = (1 + \widetilde{p}_a)\mu$ \cite{Kornilov2014Jan}.

The model \eqref{p.iamodel} becomes simplified,
\begin{equation}\label{psimple}
p_\text{simp}(\Delta t) = \widetilde{p}_a \delta(\Delta t) + (1-\widetilde{p}_a)\mu e^{-\mu \Delta t},
\end{equation}
and can be used to calculate the PDF of the n\textsuperscript{th} detection in a row, and eventually the counting statistics inside a time window $T$. Careful discussion of recovery time is needed to obtain mathematically correct results. The step-by-step derivation is presented in Appendix \ref{appendix.model}. The resulting probability $P_n$ of $n$ detections occurring within a time $T$ is

\begin{widetext}
\begin{align}	\label{methods.Pzero_final}
P_0(T) &= \frac{1-\widetilde{p}_a}{1-\widetilde{p}_a+\mu \tau_{\text{R}}}e^{-M_1},	\\
\begin{split}
P_{0<n<N}(T) &= \frac{1}{1-\widetilde{p}_a+\mu\tau_{\text{R}}} \sum_{k=0}^n \binom{n}{k} \widetilde{p}_a^{n-k} (1-\widetilde{p}_a)^k \biggl[ (k+1-\widetilde{p}_a)\mathcal{Q}_{k+1}^{n+1} - M_{n+1} \mathcal{Q}_{k}^{n+1} \\
&\quad - \Bigl( 2k+(1-k/n)\left( 1-\widetilde{p}_a \right) \Bigr)\mathcal{Q}_{k+1}^{n} + (\widetilde{p}_a k/n +2M_n)\mathcal{Q}_k^n + k\, \mathcal{Q}_{k+1}^{n-1} - (k/n + M_{n-1})\mathcal{Q}_k^{n-1} \biggr],
\end{split}
\\
\begin{split}
P_N(T) &= \frac{1}{1-\widetilde{p}_a+\mu\tau_{\text{R}}} \Biggl\{-M_0 + \sum_{k=0}^N \binom{N}{k} \widetilde{p}_a^{N-k} (1-\widetilde{p}_a)^k \biggl[ (\widetilde{p}_a k/N +2M_N)\mathcal{Q}_k^N	\\
&\qquad - \Bigl( 2k+(1-k/N)\left( 1-\widetilde{p}_a \right) \Bigr)\mathcal{Q}_{k+1}^N	+ k\, \mathcal{Q}_{k+1}^{N-1} - (k/N + M_{N-1})\mathcal{Q}_k^{N-1} \biggr]\Biggr\} + N + 1,
\end{split}
\\	\label{methods.PN+1_final}
P_{N+1}(T) &= \frac{1}{1-\widetilde{p}_a+\mu\tau_{\text{R}}} \Biggl\{ M_0 + \sum_{k=0}^{N+1} \binom{N+1}{k} \widetilde{p}_a^{N+1-k} (1-\widetilde{p}_a)^k \biggl[k \mathcal{Q}_{k+1}^{N} - \left(\frac{k}{N+1} + M_{N} \right) \mathcal{Q}_{k}^{N} \biggr] \Biggr\} - N,
\end{align}
\end{widetext}
\noindent where the terms $M$ and $\mathcal{Q}$ are defined as
\begin{align}
M_n &\coloneq \mu(T - n \tau_{\text{R}}),	\\
\mathcal{Q}_k^n &\coloneq Q_k(M_n) =
\begin{cases}
0 & \text{if}\ k=0	\\
e^{-M_n} \sum_{i=0}^{k-1}M_n^i/i! & \text{if}\  k \geq 1
\end{cases}
.
\end{align}
The number of detections where the analytical expression changes is $N = \lfloor T/\tau_{\text{R}} \rfloor$.

In the limit of $\widetilde{p}_a \to 0$, or $\widetilde{p}_a \equiv 0$ if one postulates $0^0 \coloneq 1$, the relations are reduced to the form published by M\"{u}ller for a dead-time-only process (equations (32) in ref. \cite{Muller1973Sep}).

The relations \eqref{methods.Pzero_final} to \eqref{methods.PN+1_final} are an exact model of the point process defined by eq. \eqref{psimple} and by the recovery time $\tau_{\text{R}}$. As the definition \eqref{psimple} is rather simple, the model can be conveniently verified using a Monte Carlo simulation. Any of the more complex models above cannot be expressed explicitly; the only approach then is a numerical simulation.

\section{SPAD measurements: results and discussion}

\begin{figure}
\centering
\includegraphics[width=\linewidth]{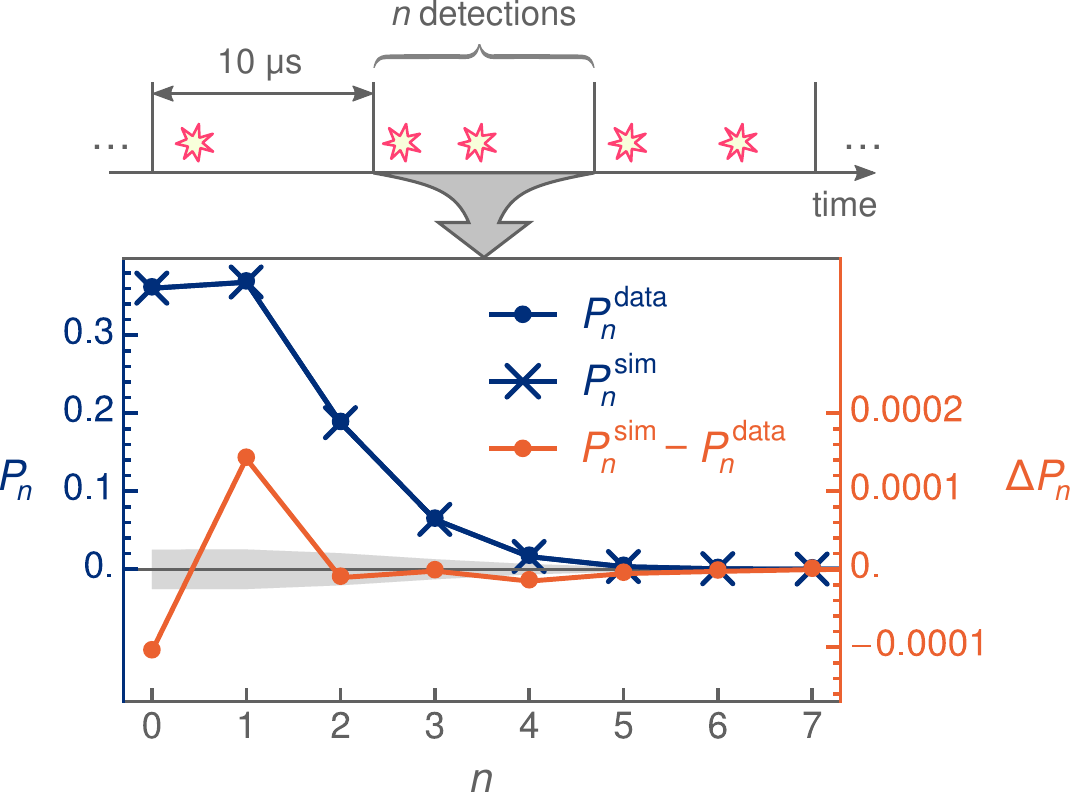}
\caption{A sample result. During a 1-hour-long measurement, the detections are binned into 10-\textmu{}s windows. The number of detections $n$ in a window then follows the probability distribution $P_n^{\text{data}}$. This is compared to a distribution $P_n^{\text{sim}}$ simulated according to \eqref{pdf.tn}. The difference between them, $\Delta P_n$, is plotted on a magnified scale (orange). Statistical error $\pm \sigma$ of $\Delta P_n$ is shown on the magnified scale as a gray area around zero. The parameters of the simulation were established using time-resolved measurements to be $\langle n_{\text{AP}}\rangle = 0.002$, $\tau_{\text{R}} = 29.1$ ns, $\alpha = p_{\text{T}}/\mu =  2$ ns. The remaining parameter $\mu$ is set so that the mean number of detections match. The full data set for multiple SPADs and count rates is given in Fig.~\ref{fig.data}.
}
\label{fig.solodata}
\end{figure}

\begin{figure*}[t]
\centering
\includegraphics[width=\linewidth]{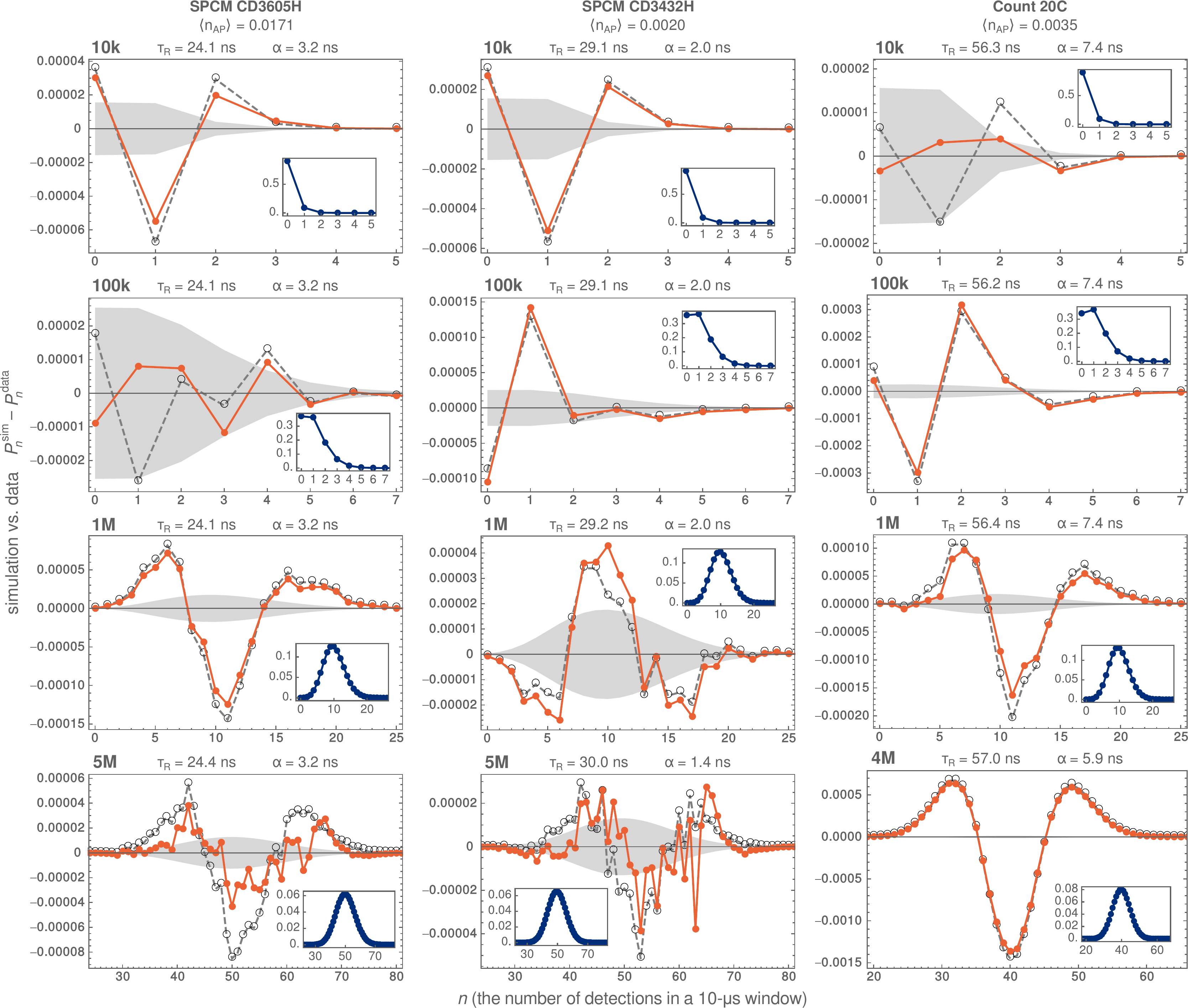}
\caption{The insets show the measured probability $P_n^{\text{data}}$ of observing $n$ counts in a 10-\textmu{}s window as a function of $n$, and the plots show the difference between the simulated probabilities $P_n^{\text{sim}}$ and the measured probabilities $P_n^{\text{data}}$. The measurements were performed using three SPAD modules (columns) and several average count rates (rows). The zero baseline represents the measured distribution and the gray area is a $\pm\sigma$ confidence interval (also see Fig.~\ref{fig.solodata}). The orange points represent a Monte Carlo simulation of the full model \eqref{pdf.tn}. The empty black points represent the analytical formulas \eqref{methods.Pzero_final} to \eqref{methods.PN+1_final}. Points within the confidence band mean a good match between the corresponding model and data, while correlation between the orange and black points mean how well the analytical model approximates the full model. Each plot shows the approximate average count rate in kilo-counts and Mega-counts per second (top left, bold), the recovery time $\tau_{\text{R}}$ and the twilight proportionality constant $\alpha=p_{\text{T}}/\mu$ (top center). The mean number of afterpulses per detection $\langle n_{\text{AP}} \rangle$ is given for each detector at the top of the column. The total variation distance for each plot is shown in Table \ref{tab.data}.
}
\label{fig.data}
\end{figure*}

We compared our predictions with the counting statistics of three different actively quenched silicon SPAD modules made by two manufacturers (Excelitas SPCM CD3605H and CD3432H, Laser Components Count 20C). First, afterpulsing was measured in a separate experiment. Each SPAD was subjected to a pulsed signal coming from an attenuated gain-switched VCSEL diode (850 nm). Using a 81-ps time-tagging module (Qutools qutau), the SPAD detection times were recorded with the photon events being announced by the electronic trigger. All subsequent events were recorded in a start-stop histogram. The pulse frequency was 478 kHz with 0.2 average detections per pulse, so afterpulsing was recorded within a 2-\textmu{}s range. This way, we established $\langle n_\text{AP} \rangle$ and the histograms served as numerical inputs of  $\nu(t)$ or $p_\text{AP}(t)$ for the simulations. The 5.5-ns dead time of the time-tagging module was shorter than the SPADs' recovery times and had no effect.

Then, we subjected the SPADs to constant continuous-wave input signals of different orders of magnitude (LED filtered at 810 nm). For the counting statistics, we chose the time window to be 10~\textmu{}s (see Fig.~\ref{fig.solodata}) in order to cover low and high mean numbers as well as saturation. The measurement time was 1 hour for each signal. Detector output was recorded by a 156-ps time-tagging module (UQDevices Logic16). Recovery time was established directly from interarrival histograms. We found out that it is not constant and apparently increases for higher rates (see Appendix \ref{appendix.recovery}). Twilight pulse probabilities were estimated from the twilight peaks in the histograms and for each detector, the constant $\alpha = p_{\text{T}}/\mu$ was established by a linear fit. Again, for higher rates, nonlinear behavior was observed for some detectors (see the values of $\alpha$ in Fig.~\ref{fig.data}). Both of these irregularities can be explained by a different thermal equilibrium in the SPAD circuit, which slightly modifies the resetting of the bias current and therefore affects twilight pulse tardiness, efficiency, and effective reset time.

For the detector CD3605H, we also had to revise the value of $\langle n_\text{AP} \rangle$, which was originally 0.0141 in the pulsed measurement, but 0.0171 in the continuous regime. This difference was observed in the respective interarrival histograms (pulsed vs. continuous 10~kcps and 100~kcps), but cannot be explained in terms of twilight pulsing. The exact cause is unknown, as there is no evidence of differences in SPAD temperature or average avalanche charge -- the factors known to affect afterpulsing probability.

\begin{table}
\centering
\caption{Total variation distance between data and model \eqref{pdf.tn}}
\begin{tabular}{rrrr}
\toprule
rate & SPCM CD3605H & SPCM CD3432H & Count 20C	\\\midrule
10k & $0.5 \times 10^{-4}$ & $0.5 \times 10^{-4}$ & $0.1 \times 10^{-4}$	\\
100k & $0.2 \times 10^{-4}$ & $1.4 \times 10^{-4}$ & $4.0 \times 10^{-4}$	\\
1M & $4.3 \times 10^{-4}$ & $1.8 \times 10^{-4}$ & $5.0 \times 10^{-4}$	\\
5M & $2.9 \times 10^{-4}$ & $2.3 \times 10^{-4}$ & (4M) $83.6 \times 10^{-4}$	\\\bottomrule
\end{tabular}
\label{tab.data}
\end{table}

The results are shown in Fig.~\ref{fig.data}. Some of the data approach the model within statistical tolerance, but most exhibit systematic errors. The differences between measured and predicted distributions are expressed by their total variation distance -- the maximum difference between probabilities of any two sets of results, $\text{TVD} = \sum_n |\Delta P_n|/2$. These are given in Table \ref{tab.data}. At this level of precision, several factors need to be considered.

First, there are small fluctuations and drifts in detection efficiency and/or LED intensity that affect the results of 1-hour-long integration. Another small contribution is a possible bistability of dark counts \cite{Karami2010}. These fluctuations would cause the measured distributions to be wider. Moreover, interarrival histograms obtained from the continuous measurements reveal that afterpulsing/twilight contributions sometimes do not behave as predicted. This could be due to SPAD temperature changing with count rate, which affects afterpulsing characteristics \cite{Stipcevic2013Oct,Anti2011Feb}. Afterpulsing is actually a doubly stochastic variable due to the avalanche current changing with each detection, but this effect is estimated to be smaller than the statistical error of our measurements. It has also been shown that afterpulsing in some thin-junction modules depends on the time of the previous detection \cite{Wayne2017}, but we have not observed any significant irregularities in this respect.

The Count 20C module diverges from the model the most. We investigated by examining the interarrival histograms of the continuous measurements (histograms of delays between two successive detections). Our model \eqref{p.stationary} predicts a negative exponential tail with an additional afterpulsing contribution for short delays (see also Fig.~3 in \cite{Humer2015Jul}). For the count rates 1~Mcps and 4~Mcps, we observed deviations from the negative exponentials that cannot be explained in terms of afterpulsing. If intensity/efficiency fluctuations are considered in the form of a mixture of negative exponentials, the result takes a convex shape on a log-scale instead of a straight exponential. However, the said data exhibit a slightly concave shape of the tail, which is incompatible with our assumptions. In theory, this could be due to the bias voltage settling too slowly so that for a few microseconds after recovery, the detection efficiency rises in the order of $\sim\!1\%$. Nevertheless, this effect contributes only slightly and the interarrival histograms alone were not sufficient to explain the counting distributions. Thus, the main source of error -- most likely of non-Markovian/non-stationary nature -- remains unknown.

Overall, some of the systematic effects could be compensated and corrected \emph{ad hoc} by fitting the parameters to each dataset individually or using parts of the interarrival histograms themselves as an input for the simulations. This would, however, be difficult to justify without additional independent measurements that would elaborate on the existing detection model. The reason is that some phenomena have similar effects on the counting distribution (mainly narrowing/widening) and na\"{i}ve data fitting would be unphysical. This is why the systematic errors were left uncorrected. Afterpulsing $\nu(t)$ was considered to be detector-specific, but unchanging; the quantities $\tau_{\text{R}}$ and $\alpha$ were observed directly from the interarrival histograms; and the intensity  $\mu$ was calculated for each data as the only free parameter to match the mean detection rate. However, none of the parameters are truly constant and more elaborate SPAD models are required to reach higher precision.

The data are also compared with the simplified analytical model \eqref{methods.Pzero_final} to \eqref{methods.PN+1_final} in Fig.~\ref{fig.data}. The approximated afterpulsing probability was chosen $\widetilde{p}_a = p_a + p_{\text{T}} - p_a p_{\text{T}}$, where $p_a = 1-e^{-\langle n_\text{AP} \rangle}$ and $p_{\text{T}} = \alpha\mu$. Fig.~\ref{fig.data} also shows that often, the approximation error of \eqref{psimple} relative to \eqref{pdf.tn} is much lower than other systematic errors. The results for the SPAD CD3605H counting statistics and mean detection rates (Figs. \ref{fig.meanrates} and \ref{fig.data}) indicate that the relative error made by the approximation alone is $\lesssim 10^{-3}$.

\section{Conclusion}

The formulation of a point process detection model allowed us to propose methods of mean-rate correction and counting statistics calculation that treat the detection aftereffects in new detail. The presented calculations can be evaluated both theoretically and using experimental data, such as numerical afterpulsing distributions. A multi-threadable algorithm was proposed that simulates the SPAD counting process. The presented methods can be used with any model of afterpulsing traps \cite{Ziarkash2018Mar}. Approximations were shown that simplify the model down to explicit formulas. Counting distributions were experimentally measured with sufficient precision to show the limitations of the proposed approach.

Our results also offer more accurate mean-rate corrections that are important for everyday rate estimation. Among applications that rely on predictable detector response are transmission measurements \cite{Sabines-Chesterking2017Jul}, single-photon imaging \cite{Ingle2019Jun}, or verification of Born's rule \cite{Kauten2017Mar}. The counting model improves the current treatments of SPAD counting statistics and its applications, such as estimating afterpulsing from a variance-to-mean ratio \cite{Tzou2015Aug}. Counting statistics also offers a new way of characterizing SPAD non-Markovian phenomena \cite{Wang2016Aug,Wayne2017}, as it is particularly affected by cumulative effects that cannot be fully distinguished in a start-stop histogram.

\appendices

\section{Point processes}
\label{appendix.point_processes}

\subsection{Preliminaries}
\label{appendix.preliminaries}

This section provides some background and more detailed discussion of point processes and their use in the main text. A one-dimensional point process on $t \in \mathbb{R}$ is a random process with a realization in a form of a set of points $\{t_i\}$. It is usually described by an intensity function $\lambda(t) \geq 0$ that represents the average density of events. The intensity can be either explicitly given or it can depend on the particular realization $\{t_i\}$ (self-exciting process). The process can be defined in term of survival probability (no events happening) between points $t_1$ and $t_2$. If we denote $\Lambda(t_1,t_2) \coloneq \int_{t_1}^{t_2} \lambda(t)\mathrm{d}t$, the survival probability is
\begin{equation} \label{PS.def}
P_S(t_1,t_2) = e^{-\Lambda(t_1,t_2)}.
\end{equation}
An infinitesimal interpretation of this is a series of narrow regions $(t,t+\mathrm{d}t)$, each having an independent probability of one event occurring equal to $\lambda(t)\mathrm{d}t$. The negative exponential is then a result of Euler's limit.

The survival probability directly leads to calculating the inter-arrival probability. We wish to calculate the probability density function (PDF) $p(t)$ of the first detection since $t_0$. First, we take the portion of events where no detection happened up to $t$, which is $P_S(t_0,t)$. Then, we wish to choose the subset where at least one detection happened during time $\mathrm{d}t$. So, we subtract the complement --  $P_S(t_0,t+\mathrm{d}t)$. The resulting probability can then be converted to PDF by $\mathrm{d}t \to 0$,
\begin{equation}
p(t) = -\frac{\partial P_S(t_0, t)}{\partial{t}} = \lambda(t) e^{-\Lambda(t_0,t)}.
\end{equation}
The ideal detection of a constant flux of photons represents the simplest case, where $\lambda(t) = \mu = \text{const.}$ The result is a well-known homogeneous Poisson process with negative-exponentially distributed inter-arrival time and a Poisson distributions of the number of events in a finite time window.

\subsection{Afterpulsing}
\label{appendix.afterpulsing}

Now let us explicitly discuss why afterpulsing is a point process under the assumption of a continuum of trap levels. We begin with a finite set of independent traps with decay rates $\{\gamma_i\}$, each having a probability to be excited $P_i$. Every trap, if excited, decays exponentially with time $\Delta t$, and the released carrier triggers the next detection. So the survival probability for one trap is a combination of not being excited or decaying later than $\Delta t$,
\begin{equation}
P_{S,i}(\Delta t)  = 1-P_i \left( 1-e^{-\gamma_i\Delta t} \right).
\end{equation}
The total survival probability is simply a product $P_S = \prod_i P_{S,i}$.

If we assume a continuum of traps with a certain excitation PDF $\rho(\gamma)$, we get $P_i = \rho(\gamma_i)\mathrm{d}\gamma$ and
\begin{align}
P_S(\Delta t) &= \lim_{\mathrm{d}\gamma \to 0} \prod_i \left[ 1 - \rho(\gamma_i)\mathrm{d}\gamma \left( 1 - e^{-\gamma_i \Delta t} \right) \right]	\\
&= \exp\left( -\int_\gamma \rho(\gamma) \left( 1 - e^{-\gamma \Delta t} \right) \mathrm{d}\gamma \right)	\\\label{PS.AP}
&= \exp\left( -\int_{0}^{\Delta t}\int_\gamma \rho(\gamma) \gamma e^{-\gamma t} \mathrm{d}\gamma \mathrm{d}t \right).
\end{align}
We can see that \eqref{PS.AP} has a form of a \emph{temporal} point process \eqref{PS.def}, where $\Lambda(\Delta t) = \int_0^{\Delta t} \rho'(t)$ and $\rho'(t) \coloneq \int \rho(\gamma) \gamma e^{-\gamma t} \mathrm{d}\gamma$. This means that each afterpulse excitation is also a point process with intensity $\rho'(t)$. Because only afterpulses at time $t>\tau_{\text{R}}$ take place, we consider the intensity $\nu(t) \coloneq \rho'(t+\tau_{\text{R}})$. The number of excited aftepulses is a Poisson-distributed variable with the mean value $\langle n_\text{AP} \rangle = \int_0^\infty \nu(t)$, typically in the order of $10^{-2}$ at most.

The aftepulsing intensity can be directly measured using a pulsed-excitation scheme and timing the detections in between the pulses. A (normalized) start-stop histogram of inter-arrival times samples the PDF
\begin{equation}
p_\text{hist}(\Delta t) = \frac{1}{1-e^{-\langle n_\text{AP} \rangle}} \nu(\Delta t) \exp\left( -\int_0^{\Delta t} \nu(t') \mathrm{d}t' \right).
\end{equation}
The limit $\langle n_\text{AP} \rangle \ll 1$ then leads to $p_\text{hist}(\Delta t) \approx \nu(\Delta t)/\langle n_\text{AP} \rangle$, which is often sufficient.

\section{Mean-rate correction}
\label{appendix.rate}

Here we cover the practical implementation of the mean detection rate calculation \eqref{rate.pointprocess}. An example code is published on CodeOcean \cite{Code}. The most essential part is obtaining the afterpulsing intensity $\nu(t)$, usually in the form of a histogram $\{H_k\}$. If the histogram bin width is $\delta t$, then $H_k \coloneq \int_{(k-1) \delta t}^{k \delta t} \nu(t)\mathrm{d}t$. We denote the corresponding values of $t_k \coloneq (k-1)\delta t$. The key step is calculating the discrete form of the cross-correlation in \eqref{f.stationary}. We use the correlation theorem for fast Fourier transforms $\mathcal{F}$ of vectors $\mathbf{x}, \mathbf{y}$, each having $N$ elements,
\begin{equation}
\FFT^{-1}\left[\FFT(\mathbf{x})\FFT(\mathbf{y})^\ast\right]_k = \sum_{i=0}^{N-1} y_i^\ast \cdot x_{(i+k)\ \mathrm{mod}\ N}.
\end{equation}
To avoid the cyclical index wrap-around in $\mathbf{x}$, both vectors' lengths can be doubled by appending zeroes to each.

Let us work with the vector $\{f_k\}_{k=1}^N$. The index offset representing recovery time would be $n_{\text{R}} = \tau_{\text{R}}/\delta t$ (rounded). Then, let
\begin{align}
\mathbf{x} &\coloneq \left\{f_{i+n_{\text{R}}}\right\}_{i=1}^{2N},	\\\label{yvector}
\mathbf{y} &\coloneq \left\{\mu\cdot\delta t + f_i \exp\left(-\mu t_i - \sum\nolimits_{j=1}^i f_j\right)\right\}_{i=1}^{2N},
\end{align}
where we consider each out-of-bound value to be zero -- namely, $t_i,f_i \coloneq 0\ \forall i>N$. Then, a single iteration step is
\begin{equation}
f_k^\text{next} = (1-p_{\text{T}})\FFT^{-1}\left[\FFT \left(\mathbf{x}\right)\FFT \left(\mathbf{y}\right)^\ast\right]_k + p_{\text{T}} f_{k+n_{\text{R}}} + H_k
\end{equation}
with the initial vector being $f_k \equiv 0$.

The next step is averaging over $\overline{p}(\Delta t)$ given in \eqref{p.stationary}, where $\overline{\lambda}_k = f_k + \mu$. Averaging beyond the value $t_N$ of $\overline{\lambda}$ needs to be written analytically. We can conveniently use the definition \eqref{yvector} with the resulting vector $f$ and write
\begin{align}
\langle \Delta t \rangle_{\overline{p}} &= (1-p_{\text{T}}) \left( \sum_i t_i y_i + \left(\frac{1}{\mu}+t_N \right)e^{-t_N \mu - \sum_i f_i} \right),	\\
\mu_\text{det} &= \left( \langle \Delta t \rangle_{\overline{p}} + \tau_{\text{R}} \right)^{-1}.
\end{align}

The main caveats here are sampling and the number of iterations. The bin width $\delta t$ needs to be short enough to neglect rounding errors in $n_{\text{R}}$, and $t_N$ needs to be long enough for $f$ to approach zero at the end, which is mainly determined by $H$. It is possible to work with a smaller bin width than the one given by the afterpulsing histogram, but then $H$ needs to be interlaced by zeroes.

\section{Derivation of the analytical counting model}
\label{appendix.model}

Here we derive the equations \eqref{methods.Pzero_final} to \eqref{methods.PN+1_final} in the main text. The process is given by the probability density function (PDF) of the time $t$ between the end of detector recovery and the next detection, where $\tau_{\text{R}}$ is a constant recovery time,
\begin{equation}\label{pinter}
p_\text{inter}(t) = \widetilde{p}_a \delta(t) + (1-\widetilde{p}_a)\mu e^{-\mu t}, \quad t \geq 0.
\end{equation}
The parameter $\mu$ is the constant temporal density, $\widetilde{p}_a$ is the afterpulse probability, and $\delta(t)$ is the Dirac delta distribution. Let us work with the temporal PDFs of the 1\textsuperscript{st}, 2\textsuperscript{nd}, \dots, $n$\textsuperscript{th} detection. First, let us consider the case when the detector is free (not recovering) at time zero. The probability of no detection up to time $t$ is $P_0^\text{free}(t) = \exp(-\mu t)$. The PDF of the first detection is simply $p_1^\text{free}(t_1) = \mu \exp(-\mu t_1)$. Then, recovery time follows, so that the time of the second detection $t_2 \geq t_1 + \tau_{\text{R}}$. The PDF of the second detection integrates over all possible times $t_1$ of the first detection:
\begin{equation}
p_2^\text{free}(t_2) = \int_0^{t_2-\tau_{\text{R}}} p_1^\text{free}(t_1) p_\text{inter}(t_2-(t_1+\tau_{\text{R}})) \text{d}t_1, \quad t_2 \geq \tau_{\text{R}}.
\end{equation}
By extension, the PDF of each detection is always a convolution of the PDF of the previous detection and PDF of the interarrival time,
\begin{align}
p_n^\text{free}(t) &= \int_{(n-2)\tau_{\text{R}}}^{t-\tau_{\text{R}}} p_{n-1}^\text{free}(t') p_\text{inter}(t-t'-\tau_{\text{R}}) \text{d}t',	\\\label{pnfree}
p_n^\text{free}(t) &= \mu e^{-\mu[t-(n-1)\tau_{\text{R}}]}	\\\nonumber
&\quad\times \sum_{k=0}^{n-1} \binom{n-1}{k} \widetilde{p}_a^{n-1-k} (1-\widetilde{p}_a)^k \mu^k \frac{\left[t-(n-1)\tau_{\text{R}}\right]^k}{k!} ,
\end{align}
where $t \geq (n-1)\tau_{\text{R}}$. Now, let us consider the probability of $n$ detections in a time window between zero and $T$. This means that the $n$\textsuperscript{th} detection happens at time $t<T$ and no more detections happen afterwards. This must be split into two cases. In the first case, the $n$\textsuperscript{th} recovery time goes beyond the time window, $t + \tau_{\text{R}} > T$. Then, no further detections inside the interval can take place. In the other case, if $t \leq T - \tau_{\text{R}}$, then the probability of no further detections occurring is the product of no afterpulsing and no detections afterwards, which is equal to  $(1-\widetilde{p}_a)\exp(-\mu(T-\tau_{\text{R}}-t))$. Both cases are possible if $n \leq \lfloor T/\tau_{\text{R}} \rfloor - 1$. If we denote the maximum amount of recoveries that fit inside the detection window $N \coloneq \lfloor T/\tau_{\text{R}} \rfloor$, then the maximum amount of detections is $N+1$. Considering the time requirements of both cases, the probability of $n$ detections is
\begin{align}
\begin{split}\label{Pnfree_general}
P_{n \leq N}^\text{free}(T) &= (1-\widetilde{p}_a) \int_{(n-1)\tau_{\text{R}}}^{T-\tau_{\text{R}}} p_n^\text{free}(t)e^{-\mu(T-\tau_{\text{R}}-t)}\,\text{d}t	\\
&\quad + \int_{T-\tau_{\text{R}}}^T \hspace{-12pt} p_n^\text{free}(t)\,\text{d}t,
\end{split}
\\[10pt]
\begin{split}\label{Pnfree_N+1}
P_{N+1}^\text{free}(T) &= \int_{N \tau_{\text{R}}}^T p_{N+1}^\text{free}(t)\,\text{d}t.
\end{split}
\end{align}
Let us now abbreviate $M_n \coloneq \mu(T-n\tau_{\text{R}})$, which could be interpreted as an ideal mean number of detections in a time window reduced by $n$ recovery times. Also, let $Q_k(x) = \exp(-x) \sum_{m=0}^{k-1}x^m/m!$ be the regularized upper incomplete Gamma function, which in this special case of $k \in \mathbb{N}^0$ represents the probability of a Poissonian variable with mean $x$ to be less than $k$ (note that $Q_0(x)=0$). Using this notation, let us substitute \eqref{pnfree} into \eqref{Pnfree_general} and \eqref{Pnfree_N+1} to obtain
\begin{align}
P_0^\text{free}(T) &= e^{-M_0},	\\\label{methods.Pnfree}
P_{1\leq n \leq N}^\text{free}(T) &= \sum_{k=0}^{n-1} \binom{n-1}{k} \widetilde{p}_a^{n-1-k} (1-\widetilde{p}_a)^k \bigg[ Q_{k+1}(M_n)	\\\nonumber
& \qquad  - Q_{k+1}(M_{n-1}) + (1-\widetilde{p}_a)\frac{M_n^{k+1}}{(k+1)!}e^{-M_n}\bigg],	\\
P_{N+1}^\text{free}(T) &= 1 - \sum_{k=0}^{N} \binom{N}{k} \widetilde{p}_a^{N-k} (1-\widetilde{p}_a)^k Q_{k+1}(M_N)
\end{align}
where the terms $Q_k(M_n)$ can be viewed with regard to the interpretations mentioned above.
These equations give a counting model assuming the detector is free at the beginning. However, if the detector is recovering at $t=0$ and keeps inactive for a certain initial time $\tau_{\text{i}}<\tau_{\text{R}}$, then the initial detection has the PDF
\begin{equation}
p_1^\text{rec}(t_1,\tau_{\text{i}}) = \widetilde{p}_a \delta(t_1-\tau_{\text{i}}) + (1-\widetilde{p}_a) \mu e^{-\mu (t_1-\tau_{\text{i}})},\hspace{5pt} t_1 \geq \tau_{\text{i}}.
\end{equation}
Like before, multiple convolutions result in the $n$\textsuperscript{th} detection PDF
\begin{align}\nonumber
p_n^\text{rec}(t,\tau_{\text{i}}) &= \widetilde{p}_a^n \delta \left(t-(n-1)\tau_{\text{R}}-\tau_{\text{i}} \right) + e^{-\mu[t-(n-1)\tau_{\text{R}}-\tau_{\text{i}}]}	\\
&\times \sum_{k=1}^n \binom{n}{k} \widetilde{p}_a^{n-k} (1-\widetilde{p}_a)^k \mu^k \frac{\left[t-(n-1)\tau_{\text{R}}-\tau_{\text{i}}\right]^{k-1}}{(k-1)!},
\end{align}
where $t \geq (n-1)\tau_{\text{R}} + \tau_{\text{i}}$. By integration analogous to \eqref{Pnfree_general}, the probability of $n$ detections in a time window $T$ then is
\begin{align}
P_0^\text{rec}(T, \tau_{\text{i}}) &= (1-\widetilde{p}_a) e^{-\mu(T-\tau_{\text{i}})},	\\\label{methods.Pnrec}
P_{1 \leq n < N}^\text{rec}(T, \tau_{\text{i}}) &= (1-\widetilde{p}_a) \widetilde{p}_a^n e^{-\mu(T-n \tau_{\text{R}} - \tau_{\text{i}})}	\\\nonumber
& + (1-\widetilde{p}_a) e^{-\mu(T-n\tau_{\text{R}}-\tau_{\text{i}})} 	\\\nonumber
&\quad\times \sum_{k=1}^n \binom{n}{k} \widetilde{p}_a^{n-k} (1-\widetilde{p}_a)^k \mu^k \frac{(T-n \tau_{\text{R}} - \tau_{\text{i}})^k}{k!}	\\\nonumber
&+ \sum_{k=1}^n \binom{n}{k} \widetilde{p}_a^{n-k} (1-\widetilde{p}_a)^k 	\\\nonumber
&\quad\times \big[ Q_k(\mu(T-n \tau_{\text{R}} - \tau_{\text{i}}))	\\\nonumber
&\qquad - Q_k(\mu(T-(n-1) \tau_{\text{R}} - \tau_{\text{i}})) \big],
\end{align}
which is similar to \eqref{methods.Pnfree}, except for the first term and the $\tau_{\text{i}}$ contribution. The first term is kept separate intentionally for consistent analytic integration in the subsequent step.

For the remaining cases of $n \geq N$, the initial time $\tau_{\text{i}}$ determines whether the $N$\textsuperscript{th} recovery time can possibly be inside the detection window or not. The border value is $\widetilde{\tau}_{\text{i}} = T - N \tau_{\text{R}}$. Therefore we need to split the two cases, while for $n=N+1$ the final recovery time always goes beyond the detection window and no more detections are possible.

\begin{align}\label{PNrec_first}
P_N^\text{rec}(T,\tau_{\text{i}} < \widetilde\tau_{\text{i}}) &= P_{1 \leq n < N}^\text{rec}(T, \tau_{\text{i}}) \Big\vert_{n = N},	\\
P_N^\text{rec}(T,\tau_{\text{i}} > \widetilde\tau_{\text{i}}) &= \int_{(N-1)\tau_{\text{R}}+\tau_{\text{i}}}^T p_N^\text{rec}(t,\tau_{\text{i}})\,\text{d}t,	\\
P_{N+1}^\text{rec}(T,\tau_{\text{i}} < \widetilde\tau_{\text{i}}) &= \int_{N\tau_{\text{R}}+\tau_{\text{i}}}^T p_{N+1}^\text{rec}(t,\tau_{\text{i}})\,\text{d}t,	\\\label{PNrec_last}
P_{N+1}^\text{rec}(T,\tau_{\text{i}} > \widetilde\tau_{\text{i}}) &= 0.
\end{align}

Now we have obtained both distributions $P_n^\text{free}(T)$ and $P_n^\text{rec}(T,\tau_{\text{i}})$ separately, where the distinction is the state of the detector at the beginning of the time window. We need to combine these cases by determining their statistical representation in a long measurement. Let us note that detection intervals are periodically distributed with a fixed length $T$, while detections follow the probabilistic point process \eqref{pinter}. So, in a long measurement, these two become uncorrelated and one can assume that the distribution of window beginnings with respect to detection events is completely random. Therefore, the proportion of ``free'' windows to ``rec'' windows is equal to the proportion of overall times when the detector was free and blocked, respectively, $\langle t \rangle : \tau_{\text{R}}$. Additionally, the distribution of $\tau_{\text{i}}$ is uniform between zero and $\tau_{\text{R}}$. Taking both of these into account, the overall probability of $n$ detections becomes a mixture
\begin{align}
P_n(T) &= \frac{\langle t \rangle}{\langle t \rangle + \tau_{\text{R}}} P_n^\text{free}(T)	\\\nonumber
& \quad + \left( 1 - \frac{\langle t \rangle}{\langle t \rangle + \tau_{\text{R}}} \right) \frac{1}{\tau_{\text{R}}} \int_0^{\tau_{\text{R}}} P_n^\text{rec}(T,\tau_{\text{i}})\,\text{d}\tau_{\text{i}}.
\end{align}
This mixture needs to be evaluated separately for the cases of $n=0$, $n<N$, $n=N$, and $n=N+1$, because the probability distributions differ and the integration over $\tau_{\text{i}}$ needs to be split to accommodate the piecewise definitions \eqref{PNrec_first} to \eqref{PNrec_last}.
After integration, renumbering of the summation indices and using the property $Q_0(x)=0$, we obtain the equations \eqref{methods.Pzero_final} to \eqref{methods.PN+1_final} in the main text.

\section{Numerical simulation}
\label{appendix.simulation}

Here we introduce an algorithm that generates the process defined by \eqref{pdf.tn} to simulate the SPAD operation \cite{Code}. Its output is a histogram of the number of detections in a time window $T$, which upon normalization yields $P_n^{\text{sim}}$ (see Figs. \ref{fig.solodata} and \ref{fig.data} in the main text).

Below is the pseudo-code of one cycle of the simulation. It updates the variable \texttt{time} to the time of the next detection event. The detection event counter \texttt{detectionEvents} is incremented and if the end of the current time window is reached, the count histogram is updated.

At the beginning, \texttt{time} holds the time of the previous detection event. The array \texttt{AP\_queue} stores the time of decay of all currently populated afterpulsing traps. After the recovery time is added, all the afterpulses that happened during recovery are removed from the queue. The number of new afterpulses \texttt{nAP} is then generated as a Poisson variable. \texttt{random(0,1)} gives a number uniformly distributed between 0 and 1, \texttt{AP\_MEAN} is $\langle n_{\text{AP}} \rangle$, and \texttt{PoissonInvCDF} is the quantile function (inverse cumulative distribution) of the Poisson distribution. Each afterpulse is assigned a detection time by \texttt{AP()}, which is a quantile function of the afterpulsing temporal distribution $p_{\text{AP}}(t) \approx \nu(t)/\langle n_{\text{AP}} \rangle$. Next, if a twilight pulse happens with a probability \texttt{p\_twilight}, no additional time is added and the detection happens right after recovery time. Otherwise, the time of the next photon absorption is calculated as an exponentially distributed variable. Then, the afterpulsing queue is searched and if there is an afterpulse happening earlier, the detection time is updated. In the next part, if the detection happens in a new time window ($T=$~\texttt{WINDOW}), the counting histogram is updated and the counter reset. The beginning of the time window is kept at zero to avoid large floating-point values, and so all time values are offset. Finally, the detection counter is incremented.

\noindent
A single loop cycle is
\begin{Verbatim}[samepage,commandchars=\\\{\}]
time += recoveryTime
\textcolor{gray}{// remove old afterpulses}
for (time_AP in AP_queue)
	if (time_AP <= time)
		RemoveFromQueue(time_AP)
\textcolor{gray}{// add new afterpulses}
nAP = PoissonInvCDF(AP_MEAN, random(0,1))
for (i=0; i<nAP; i++)
	time_AP = time + AP(random(0,1))
	AddToQueue(time_AP)
if (random(0,1) > p_twilight)
    \textcolor{gray}{// photon absorption}
    time =  time - log(random(0,1))/rate
    \textcolor{gray}{// checking for afterpulses}
    for (time_AP in AP_queue)
    	if (time_AP < time)
    		time = time_AP
end
\textcolor{gray}{// Now `time' holds the arrival time}
\textcolor{gray}{// of the detection to be recorded.}
while (time > WINDOW)
    time -= WINDOW
    for (time_AP in AP_queue)
    	time_AP -= WINDOW
    incrementHistogram(detectionEvents)
    detectionEvents = 0
end
detectionEvents += 1
\end{Verbatim}

The above loop cycle provides one sample of a detection event. It is written to simulate the process established by \eqref{pdf.tn}, but it can be simplified or expanded depending on how complex the detection model is. It can be conveniently run in multiple threads. An important technical note is that within one loop, each instance of a random number \texttt{random(0,1)} should come from a separate pseudo-random-number generator. In our implementation, we found that if this condition is not met, the insufficiency in randomness is statistically observable in the simulation.

The number of runs for verifying of relations \eqref{methods.Pzero_final} to \eqref{methods.PN+1_final} was $10^{11}$. The same number of runs was used to verify the precision of the iterative mean-detection-rate formula \eqref{rate.pointprocess} and it was found accurate within the statistical precision $\sigma=3\times10^{-6}\mu$. An implementation of this algorithm is published on CodeOcean and GitHub \cite{Code}.

\clearpage

\section{Recovery time}
\label{appendix.recovery}

\begin{figure}[h]
\centering
\includegraphics[width=.85\linewidth]{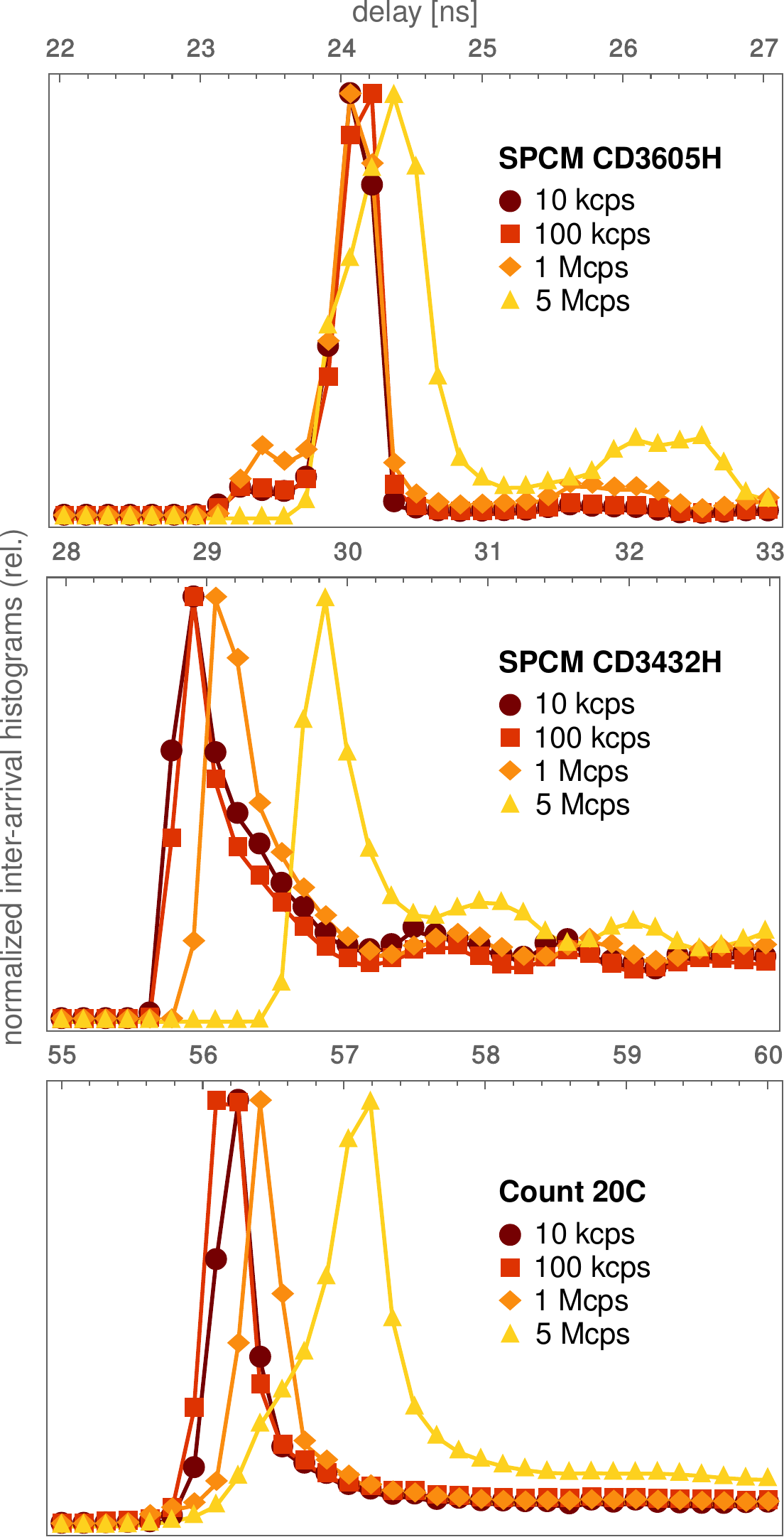}
\caption{The recovery time histograms for all three tested SPADs. Each color represents a certain count rate. All the points on the left of the peaks are zero.}
\label{fig.recoverytimes}
\end{figure}

Here we show how the SPADs exhibit changes in recovery time. Fig. \ref{fig.recoverytimes} shows the histograms of delays between two successive detections. The histograms are scaled so that the twilight/afterpulsing peak locations can be distinguished. The peaks mark the earliest detections and determine recovery time $\tau_{\text{R}}$. Generally, $\tau_{\text{R}}$ increases with rate and the changes become significant as the detectors starts being saturated.

\IEEEtriggeratref{26}



\begin{thebibliography}{10}
\providecommand{\url}[1]{#1}
\csname url@samestyle\endcsname
\providecommand{\newblock}{\relax}
\providecommand{\bibinfo}[2]{#2}
\providecommand{\BIBentrySTDinterwordspacing}{\spaceskip=0pt\relax}
\providecommand{\BIBentryALTinterwordstretchfactor}{4}
\providecommand{\BIBentryALTinterwordspacing}{\spaceskip=\fontdimen2\font plus
\BIBentryALTinterwordstretchfactor\fontdimen3\font minus
  \fontdimen4\font\relax}
\providecommand{\BIBforeignlanguage}[2]{{%
\expandafter\ifx\csname l@#1\endcsname\relax
\typeout{** WARNING: IEEEtran.bst: No hyphenation pattern has been}%
\typeout{** loaded for the language `#1'. Using the pattern for}%
\typeout{** the default language instead.}%
\else
\language=\csname l@#1\endcsname
\fi
#2}}
\providecommand{\BIBdecl}{\relax}
\BIBdecl

\bibitem{Migdall2013Book}
A.~Migdall, S.~V. Polyakov, J.~Fan, and J.~C. Bienfang, \emph{Single-Photon
  Generation and Detection}.\hskip 1em plus 0.5em minus 0.4em\relax Academic
  Press, Dec 2013.

\bibitem{Chunnilall2014Jul}
C.~J. Chunnilall, I.~P. Degiovanni, S.~K\"{u}ck, I.~M\"{u}ller, and A.~G. Sinclair,
  ``Metrology of single-photon sources and detectors: a review,'' \emph{Optical
  Engineering}, vol.~53, no.~8, p. 081910, Jul 2014.

\bibitem{Buller2007Dec}
G.~S. Buller, R.~E. Warburton, S.~Pellegrini, J.~S. Ng, J.~P.~R. David,
  L.~J.~J. Tan, A.~B. Krysa, and S.~Cova, ``{Single-photon avalanche diode
  detectors for quantum key distribution},'' \emph{IET Optoelectron.}, vol.~1,
  no.~6, pp. 249--254, Dec 2007.

\bibitem{Eraerds2007Oct}
P.~Eraerds, M.~Legr\'{e}, A.~Rochas, H.~Zbinden, and N.~Gisin, ``{SiPM} for fast
  photon-counting and multiphoton detection,'' \emph{Optics Express}, vol.~15,
  no.~22, pp. 14\,539--14\,549, Oct 2007.

\bibitem{Chesi2019May}
G.~Chesi, L.~Malinverno, A.~Allevi, R.~Santoro, M.~Caccia, A.~Martemiyanov, and
  M.~Bondani, ``{Optimizing Silicon photomultipliers for Quantum Optics},''
  \emph{Sci. Rep.}, vol.~9, no. 7433, pp. 1--12, May 2019.

\bibitem{Zhang2018Feb}
L.~Zhang, D.~Chitnis, H.~Chun, S.~Rajbhandari, G.~Faulkner, D.~O'Brien, and
  S.~Collins, ``{A Comparison of APD- and SPAD-Based Receivers for Visible
  Light Communications},'' \emph{J. Lightwave Technol.}, vol.~36, no.~12, pp.
  2435--2442, Feb 2018.

\bibitem{Bruschini2019Sep}
C.~Bruschini, H.~Homulle, I.~M. Antolovic, S.~Burri, and E.~Charbon,
  ``{Single-photon avalanche diode imagers in biophotonics: review and
  outlook},'' \emph{Light Sci. Appl.}, vol.~8, no.~87, pp. 1--28, Sep 2019.

\bibitem{Kalashnikov2012Jul}
D.~A. Kalashnikov, S.-H. Tan, T.~{\relax Sh}. Iskhakov, M.~V. Chekhova, and
  L.~A. Krivitsky, ``{Measurement of two-mode squeezing with photon number
  resolving multipixel detectors},'' \emph{Opt. Lett.}, vol.~37, no.~14, pp.
  2829--2831, Jul 2012.

\bibitem{Chesi2019Mar}
G.~Chesi, L.~Malinverno, A.~Allevi, R.~Santoro, M.~Caccia, and M.~Bondani,
  ``{Measuring nonclassicality with silicon photomultipliers},'' \emph{Opt.
  Lett.}, vol.~44, no.~6, pp. 1371--1374, Mar 2019.

\bibitem{Lubin2019Nov}
G.~Lubin, R.~Tenne, I.~Michel~Antolovic, E.~Charbon, C.~Bruschini, and D.~Oron,
  ``{Quantum correlation measurement with single photon avalanche diode
  arrays},'' \emph{Opt. Express}, vol.~27, no.~23, pp. 32\,863--32\,882, Nov
  2019.

\bibitem{Antolovic2018Aug}
I.~M. Antolovic, C.~Bruschini, and E.~Charbon, ``{Dynamic range extension for
  photon counting arrays},'' \emph{Opt. Express}, vol.~26, no.~17, pp.
  22\,234--22\,248, Aug 2018.

\bibitem{Brida2009Nov}
G.~Brida, I.~P. Degiovanni, F.~Piacentini, V.~Schettini, S.~V. Polyakov, and
  A.~Migdall, ``{Scalable multiplexed detector system for high-rate
  telecom-band single-photon detection},'' \emph{Rev. Sci. Instrum.}, vol.~80,
  no.~11, p. 116103, Nov 2009.

\bibitem{Polyakov2007Feb}
S.~V. Polyakov and A.~L. Migdall, ``High accuracy verification of a
  correlated-photon-based method for determining photon-counting detection
  efficiency,'' \emph{Optics Express}, vol.~15, no.~4, pp. 1390--1407, Feb
  2007.

\bibitem{Cohen2018Jul}
L.~Cohen, Y.~Pilnyak, D.~Istrati, N.~M. Studer, J.~P. Dowling, and H.~S.
  Eisenberg, ``Absolute calibration of single-photon and multiplexed
  photon-number-resolving detectors,'' \emph{Physical Review A}, vol.~98,
  no.~1, p. 013811, Jul 2018.

\bibitem{Cova1996Apr}
S.~Cova, M.~Ghioni, A.~Lacaita, C.~Samori, and F.~Zappa, ``Avalanche
  photodiodes and quenching circuits for single-photon detection,''
  \emph{Applied Optics}, vol.~35, no.~12, pp. 1956--1976, Apr 1996.

\bibitem{Cheng2016Mar}
Z.~Cheng, X.~Zheng, D.~Palubiak, M.~J. Deen, and H.~Peng, ``{A Comprehensive
  and Accurate Analytical SPAD Model for Circuit Simulation},'' \emph{IEEE
  Trans. Electron Devices}, vol.~63, no.~5, pp. 1940--1948, Mar 2016.

\bibitem{Ware2007Jan}
M.~Ware, A.~Migdall, J.~C. Bienfang, and S.~V. Polyakov, ``Calibrating
  photon-counting detectors to high accuracy: background and deadtime issues,''
  \emph{Journal of Modern Optics}, vol.~54, no. 2-3, pp. 361--372, Jan 2007.

\bibitem{Cova1991Dec}
S.~Cova, A.~Lacaita, and G.~Ripamonti, ``{Trapping phenomena in avalanche
  photodiodes on nanosecond scale},'' \emph{IEEE Electron Device Letters},
  vol.~12, no.~12, pp. 685--687, Dec 1991.

\bibitem{Humer2015Jul}
G.~Humer, M.~Peev, C.~Schaeff, S.~Ramelow, M.~Stip\v{c}evi\'{c}, and R.~Ursin, ``A
  simple and robust method for estimating afterpulsing in single photon
  detectors,'' \emph{Journal of Lightwave Technology}, vol.~33, no.~14, pp.
  3098--3107, Jul 2015.

\bibitem{Ziarkash2018Mar}
A.~W. Ziarkash, S.~K. Joshi, M.~Stip\v{c}evi\'{c}, and R.~Ursin, ``Comparative study
  of afterpulsing behavior and models in single photon counting avalanche photo
  diode detectors,'' \emph{Scientific Reports}, vol.~8, no.~1, p. 5076, Mar
  2018.

\bibitem{Zappa2009Aug}
F.~Zappa, A.~Tosi, A.~D. Mora, and S.~Tisa, ``{SPICE modeling of single photon
  avalanche diodes},'' \emph{Sens. Actuators, A}, vol. 153, no.~2, pp.
  197--204, Aug 2009.

\bibitem{Code}
\BIBentryALTinterwordspacing
I.~Straka, ``{SPAD} counting model,'' 2020. [Online]. Available:\\
  \url{https://doi.org/10.24433/CO.8487128.v1},\\
  \url{https://github.com/ivo-s/SPAD-counting-model}
\BIBentrySTDinterwordspacing

\bibitem{Muller1973Sep}
J.~W. M\"{u}ller, ``Dead-time problems,'' \emph{Nuclear Instruments and Methods},
  vol. 112, no.~1, pp. 47--57, Sep 1973.

\bibitem{Rapp2019May}
J.~Rapp, Y.~Ma, R.~M.~A. Dawson, and V.~K. Goyal, ``{Dead Time Compensation for
  High-Flux Ranging},'' \emph{IEEE Trans. Signal Process.}, vol.~67, no.~13,
  pp. 3471--3486, May 2019.

\bibitem{Kornilov2014Jan}
V.~Kornilov, ``Effects of dead time and afterpulses in photon detector on
  measured statistics of stochastic radiation,'' \emph{Journal of the Optical
  Society of America A}, vol.~31, no.~1, pp. 7--15, Jan 2014.

\bibitem{Wang2016Aug}
F.-X. Wang, W.~Chen, Y.-P. Li, D.-Y. He, C.~Wang, Y.-G. Han, S.~Wang, Z.-Q.
  Yin, and Z.-F. Han, ``Non-{M}arkovian property of afterpulsing effect in
  single-photon avalanche detector,'' \emph{Journal of Lightwave Technology},
  vol.~34, no.~15, pp. 3610--3615, Aug 2016.

\bibitem{Snyder1991}
D.~L. Snyder and M.~I. Miller, \emph{Random Point Processes in Time and
  Space}.\hskip 1em plus 0.5em minus 0.4em\relax Springer-Verlag New York,
  1991.

\bibitem{Stipcevic2013May}
M.~Stip\v{c}evi\'{c} and D.~J. Gauthier, ``{Precise Monte Carlo simulation of
  single-photon detectors},'' \emph{Advanced Photon Counting Techniques VII},
  vol. 8727, p. 87270K, May 2013.

\bibitem{Tzou2015Aug}
B.-W. Tzou, J.-Y. Wu, Y.-S. Lee, and S.-D. Lin, ``Method to evaluate
  afterpulsing probability in single-photon avalanche diodes,'' \emph{Optics
  Letters}, vol.~40, no.~16, pp. 3774--3777, Aug 2015.

\bibitem{Horoshko2017Jan}
D.~B. Horoshko, V.~N. Chizhevsky, and S.~Y. Kilin, ``Afterpulsing model based
  on the quasi-continuous distribution of deep levels in single-photon
  avalanche diodes,'' \emph{Journal of Modern Optics}, vol.~64, no.~2, pp.
  191--195, Jan 2017.

\bibitem{Straka2018Apr}
I.~Straka, J.~Mika, and M.~Je\v{z}ek, ``Generator of arbitrary classical photon
  statistics,'' \emph{Optics Express}, vol.~26, no.~7, pp. 8998--9010, Apr
  2018.

\bibitem{Itzler2012Oct}
M.~A. Itzler, X.~Jiang, and M.~Entwistle, ``Power law temporal dependence of
  {InGaAs}/{InP} {SPAD} afterpulsing,'' \emph{Journal of Modern Optics},
  vol.~59, no.~17, pp. 1472--1480, Oct 2012.

\bibitem{Ghioni2008Feb}
M.~Ghioni, A.~Gulinatti, I.~Rech, P.~Maccagnani, and S.~Cova, ``{Large-area
  low-jitter silicon single photon avalanche diodes},'' \emph{Quantum Sensing
  and Nanophotonic Devices V}, vol. 6900, p. 69001D, Feb 2008.

\bibitem{Karami2010}
M.~A. Karami, L.~Carrara, C.~Niclass, M.~Fishburn, and E.~Charbon, ``{RTS}
  noise characterization in single-photon avalanche diodes,'' \emph{{IEEE}
  Electron Device Letters}, vol.~31, no.~7, pp. 692--694, jul 2010.

\bibitem{Stipcevic2013Oct}
M.~Stip\v{c}evi\'{c}, D.~Wang, and R.~Ursin, ``Characterization of a commercially
  available large area, high detection efficiency single-photon avalanche
  diode,'' \emph{Journal of Lightwave Technology}, vol.~31, no.~23, pp.
  3591--3596, Oct 2013.

\bibitem{Anti2011Feb}
M.~Anti, A.~Tosi, F.~Acerbi, and F.~Zappa, ``{Modeling of afterpulsing in
  single-photon avalanche diodes},'' \emph{Physics and Simulation of
  Optoelectronic Devices XIX}, vol. 7933, p. 79331R, Feb 2011.

\bibitem{Wayne2017}
M.~A. Wayne, J.~C. Bienfang, and S.~V. Polyakov, ``Simple autocorrelation
  method for thoroughly characterizing single-photon detectors,'' \emph{Optics
  Express}, vol.~25, no.~17, p. 20352, 2017.

\bibitem{Sabines-Chesterking2017Jul}
J.~Sabines-Chesterking, R.~Whittaker, S.~K. Joshi, P.~M. Birchall, P.~A.
  Moreau, A.~McMillan, H.~V. Cable, J.~L. O{'}Brien, J.~G. Rarity, and J.~C.~F.
  Matthews, ``{Sub-Shot-Noise Transmission Measurement Enabled by Active
  Feed-Forward of Heralded Single Photons},'' \emph{Phys. Rev. Appl.}, vol.~8,
  no.~1, p. 014016, Jul 2017.

\bibitem{Ingle2019Jun}
A.~{Ingle}, A.~{Velten}, and M.~{Gupta}, ``High flux passive imaging with
  single-photon sensors,'' in \emph{2019 IEEE/CVF Conference on Computer Vision
  and Pattern Recognition (CVPR)}, June 2019, pp. 6753--6762.

\bibitem{Kauten2017Mar}
T.~Kauten, R.~Keil, T.~Kaufmann, B.~Pressl, {\v{C}}.~Brukner, and G.~Weihs,
  ``{Obtaining tight bounds on higher-order interferences with a 5-path
  interferometer},'' \emph{New J. Phys.}, vol.~19, no.~3, p. 033017, Mar 2017.

\end{thebibliography}
\end{document}